\documentclass[aps,twocolumn,showpacs,amsmath,amssymb,eqsecnum,eqnarray,array]{revtex4}
\usepackage{anysize}
\marginsize{1.9cm}{1.9cm}{1.55cm}{1.55cm}
\usepackage{graphicx}
\usepackage{dcolumn}
\usepackage{bm}
\begin{document}

\title{Why a magnetized quantum wire can act as an {\em active} laser medium\\}
\author{Manvir S. Kushwaha}
\address
{Department of Physics and Astronomy, Rice University, P.O. Box 1892, Houston, TX 77251, USA}

\date{\today}

\begin{abstract}

The fundamental issues associated with the magnetoplasmon excitations are investigated in a quantum wire
characterized by a confining harmonic potential and subjected to a perpendicular magnetic field.
Essentially, we embark on the device aspects of the intersubband collective (magnetoroton) excitation
which observes a negative group velocity between the maxon and the roton. The computation of the gain
coefficient suggests an interesting and important application: the electronic device based on such
magnetoroton modes can act as an {\it active} laser medium.

\end{abstract}
\maketitle



The current progress in nanofabrication technology and the ability to tailor potentials and interactions is
stimulated by the world-wide quest to develop exotic high-speed, low-power devices that are small enough, sharp
enough, and uniform enough to behave the way theory says they should. Ever increasing excitement behind the
research interest in the systems of reduced dimensionality can now reasonably be attributed to the discovery of
the quantum Hall effects. These are the man-made semiconductor quantum structures generally known as quantum
wells, quantum wires, and quantum dots in which the charge carriers exposed to external probes such as electric
and/or magnetic fields can exhibit unprecedented quantal effects that strongly modify their behavior [1]. In the
present work, we are concerned with the theoretical investigation of quantum wires or (more realistically) a
quasi-one dimensional electron gas (Q1DEG) originally proposed by Sakaki in 1980 [see, e.g., Ref. 2].

The proposal of semiconductor quantum wire structures was motivated by the suggestion [2] that 1D $k$-space
restriction would severely reduce the impurity scattering, thereby substantially enhancing the
low-temperature electron mobilities. Consequently, the technological promise that emerges is the route to
the faster transistors and other electronic devices fabricated out of the quantum wires. A tremendous
research interest burgeoned in quantum wires owes not only to their potential applications, but also to the
fundamental physics involved. For instance, they have offered us an excellent (unique) opportunity to study
the real 1D fermi gases in a relatively controlled manner [1]. Theoretical development of electronic, optical,
and transport phenomena in quantum wires has also been the subject of an intense controversy whether the
system is best describable as Tomonaga-Luttinger liquid or as Fermi liquid. This issue was elegantly resolved
by Das Sarma and coworkers [3] who consistently justified the use of Fermi-liquid-like theories for describing
the realistic quantum wires.

This paper aims at reporting on the extensive investigation of the charge-density excitations in a magnetized
quantum wire in a two-subband model within the framework of Bohm-Pines' random-phase approximation (RPA) [1].
The main focus of our study is the intersubband collective (magnetoroton) excitation which changes the sign of
its group velocity twice before merging with the respective single-particle continuum. By definition, a roton
is an elementary excitation whose dispersion relation shows a linear increase from the origin, but exhibits
first a maximum, and then a minimum in energy as the momentum increases. Excitations with momenta in the linear
region are called phonons; those with momenta near the maximum are called maxons; and those with momenta close
to the minimum are called rotons.

In Q1DEG, the magnetoroton mode was predicted within the framework of Hartree-Fock approximation in 1992 [4]
and it was soon verified in the resonant Raman scattering experiments [5]. A rigorous theoretical finding of
the magnetoroton (MR) modes in the realistic quantum wires had, however, been elusive until quite recently
[6-9]. After a brief theoretical background, we recall the relevant aspects of the MR mode from our recent
studies on Q1DEG [7-9], compute the gain coefficient as an evidence, and systematically argue and substantiate
that the electronic device based on such MR modes can act as an {\it active} laser medium.

The motivation behind this communication is grounded crucially in the device aspects emerging from the factual
fundamental issues hypothesized in Refs. 7, 8, and 9.


We consider a Q1DEG in the $x-y$ plane with a confining harmonic potential $V(x)=\frac{1}{2}m^*\omega_0^2 x^2$
oriented along the $x$ direction and a magnetic field $B$ applied along the $z$ direction in the Landau gauge
[$\bf {A}=(0,Bx,0)$]. Here $\omega_0$ is the characteristic frequency of the harmonic oscillator and $m^*$ the
electron effective mass of the system. The resultant system is a realistic quantum wire with free propagation
along the $y$ direction and the {\em magnetoelectric} quantization along the $x$ direction.
A realistic quantum wire as defined above is characterized by the eigenfunction
$\psi_{n}(k_y)=1/\sqrt{L_y}\, e^{ik_y y}\,\phi_n(x+x_c)$, where $\phi_n(x+x_c)$ is the Hermite function,
and the eigenenergy $\epsilon_{nk_y} = (n + \frac{1}{2})\,\hbar\tilde{\omega} + \hbar^2 k_y^2/(2 m_r)$,
where $L_y$, $n$, $x_c=k_y ({\it l}_d^4/{\it l}_c^2)$, ${\it l}_c=\sqrt{\hbar/(m^*\omega_c)}$,
${\it l}_d=\sqrt{\hbar/(m^*\tilde{\omega})}$, $m_r=m^*(\tilde{\omega}^2/\omega_0^2)$, are, respectively, the
normalization length, the hybrid magnetoelectric subband (HMES) index, the center of the cyclotron orbit
with radius ${\it l}_d$, the magnetic length, the effective magnetic length, and the renormalized effective
mass. Here the hybrid frequency $\tilde{\omega}=\sqrt{\omega_c^2+\omega_0^2}$. The effective magnetic length
${\it l}_d$ refers to the typical width of the wave function and reduces to the magnetic length ${\it l}_c$
if the confining potential is zero (i.e., if $\omega_0=0$). In the limit of a strong magnetic field, the
renormalized mass $m_r$ becomes infinite and the system undergoes a cross-over to the 2DES and hence the
Landau degeneracy is recovered.


For the illustrative numerical examples, we have been focusing  on the {\em narrow} channels of the narrow-gap
In$_{1-x}$Ga$_x$As system. The material parameters used are: effective mass $m^*=0.042 m_{_0}$, the background dielectric constant $\epsilon_{_b}=13.9$, 1D charge density $n_{1D}=1.0\times 10^{6}$ cm$^{-1}$, confinement
energy $\hbar\omega_0=2.0$ meV, and the effective confinement width of the harmonic potential well, estimated
from the extent of the Hermite function, $w_{eff}=40.19$ nm. Notice that the Fermi energy $\epsilon_F$ varies
in the case where the charge density ($n_{1D}$), the magnetic field ($B$), or the confining potential ($\hbar \omega_{0}$) is varied. Thus we set out on the extensive investigation associated with the magnetoplasmon
excitations in a Q1DEG subjected to a perpendicular magnetic field $B$ at T=0 K [7-9].

The full-fledged MR mode was observed in the similar physical conditions in the magnetized quantum wires made up
of narrow-gap In$_{1-x}$Ga$_x$As systems in a rather different context [6]. Subsequently, we extensively studied
the dependence of its propagation characteristics on the charge-density, confinement potential, and magnetic
field [7]. Since the existence of this MR mode is exclusively attributed to the applied (perpendicular) magnetic
field, we also scrutinized its group velocity for numerous values of the magnetic field in order to judge when
and where this otherwise regular intersubband mode starts taking up the magnetoroton character. It was found that
there really is a minimum (threshold) value of the magnetic field ($B_{th}$) below which this MR does not exist
and a regular intersubband magnetoplasmon survives. The $B_{th}$, for the present system, is defined as $B_{th}
\simeq 1.0$ T. There it was also established that both maxon and roton are the higher density of excitation states.

Later, we also embarked on investigating the inverse dielectric functions (IDF) for this system under the similar physical conditions [8]. The motivation there was not solely to reaffirm the fact that the poles of the IDF and
the zeros of the dielectric function (DF) yield exactly identical excitation spectrum, but also to pinpoint
the advantage of the former over the latter. For instance, the imaginary (real) part of the IDF sets to furnish a significant measure of the longitudinal (Hall) resistance in the system. Moreover, the quantity
Im [$\epsilon^{-1} (q,\omega)$] also implicitly provides the reasonable estimates of the inelastic electron
(or Raman) scattering cross-section $S$($q$) for a given system. For the details of the formalism of the IDFs for
the quasi-n dimensional systems, a reader is referred to Ref. 11.

The main concern here is the MR mode between maxon and roton where it exhibits negative group velocity (NGV). We
believe that the occurrence of NGV is quite unusual and hence must have some dramatic consequences. It turns out
that the NGV matters -- in the striking phenomena such as techyon-like (superluminal) behavior [10], anomalous
dispersion in the gain medium, a state with inverted population (likely) characterized by negative temperature, a
medium having the ability of amplifying a small optical signal and hence serving as an {\em active} laser medium, ...etc. [7]. We focus on the latter and choose to compute the gain coefficient $\alpha$($\omega$) in order to materialize the notion of a quantum wire acting as an {\em active} laser medium.

\begin{figure}[htbp]
\includegraphics*[width=8cm,height=9cm]{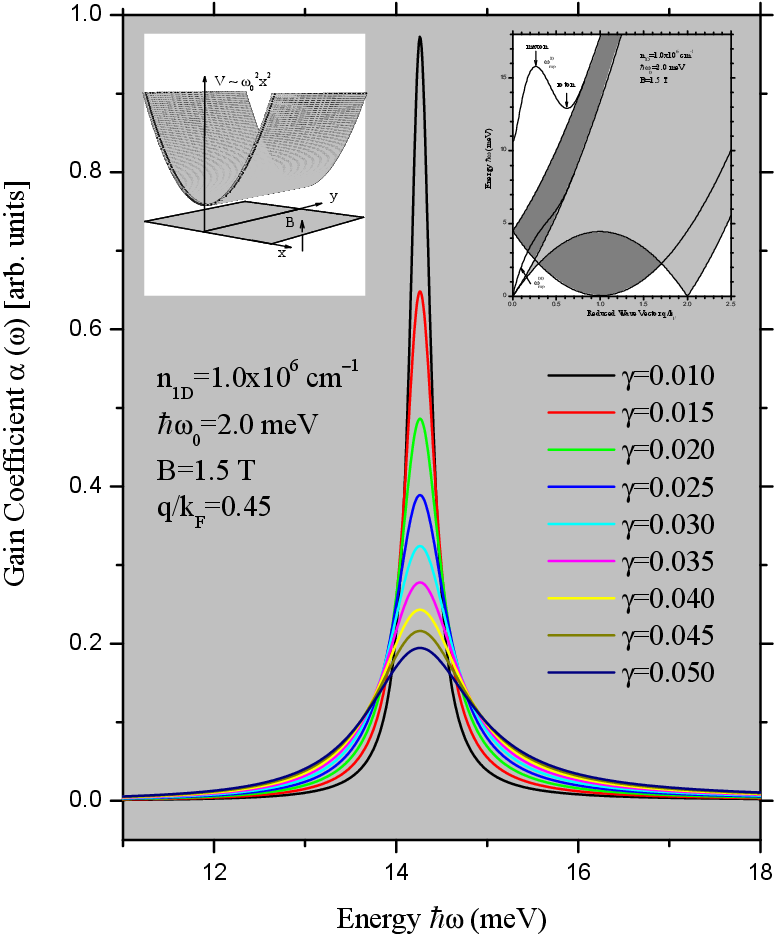}
\caption{(Color online) The gain coefficient $\alpha$($\omega$) as a function of the excitation energy
$\hbar\omega$ for several values of the damping factor and for the given values of the magnetic field $B$,
the charge density $n_{1d}$, and the confinement energy $\hbar\omega_0$. The parameters are as given inside
the picture. The inset on the top-left shows the model quantum wire investigated here. The inset on the
top-right illustrates the magnetoplasmon excitation spectrum in a two-subband model within the full RPA [8].
The crux of the matter here is the intersubband magnetoplasmon (or magnetoroton) mode which observes one
maximum [the maxon] and one minimum [the roton] (after starting at $q=0$ and $\hbar \omega =10.65$ meV)
before merging with the respective single-particle continuum.}
\label{fig1}
\end{figure}

In the classical electrodynamics, it makes sense to express $\alpha$($\omega$) in terms of Im [$\chi (\omega)$],
with $\chi$ being the susceptibility. But the story takes a different turn when it comes to the quantum systems,
as is the case here. Despite the fact that it is the $\chi$ that, generally, gives rise to the (mathematically)
complex nature of the DF or the IDF, we must recognize that $\chi$ contains only the single-particle response,
whereas the IDF can provide both single-particle and collective responses. Since our concern here is the MR which happens to be the intersubband collective (magnetoplasmon) excitation, we ought to search $\alpha$($\omega$) in
terms of Im [$\epsilon^{-1}_{inter} (q, \omega)$] rather than Im [$\chi_{inter} (q, \omega)$].

Figure 1 illustrates the computation of the gain coefficient $\alpha$($\omega$) as a function of the excitation
energy $\hbar\omega$ for various values of the damping factor $\gamma$. The gain coefficient in the context of
the laser amplification is defined as $\alpha$($\omega$)= ($\omega/2c$) Im [$\epsilon^{-1}_{inter} (q,\omega)$];
$c$ being the speed of light in vacuum. The symbol Im refers to the imaginary part of nonlocal, dynamic, inverse
dielectric function (considered only for the relevant intersubband [or magnetoroton] excitations). The gain
coefficient that persists due to the electronic transitions shows a maximum at $\hbar\omega \simeq 14.26$ meV for
the damping factor $\gamma=0.010$. It is not just by chance that the peak position occurs at the expected values
of ($q, \omega$) in the excitation spectrum. As it is intuitively expected, the peak turns towards the lower
energy with increasing damping factor. It is reasonable that an amplifier device such as a laser gain medium
cannot maintain a fixed gain for arbitrarily high input powers, because this would require adding arbitrary
amounts of power to the amplified signal. Therefore, the gain must be reduced for high input powers; this
phenomenon is called gain saturation. In the case of a laser gain medium, it is widely known that the gain does
not instantly adjust to the level according to the optical input power, because the gain medium stores some
amount of energy, and the stored energy determines the gain.

Given the sequence of instances manifesting in the system, it makes sense to reason that the applied magnetic
field drives the system to the metastable (or non-equilibrium) state [12] that gives rise to the population
inversion so that the gain rather than absorption occurs at the frequencies of interest. This is attested by the existing negative group velocity (NGV) associated with the anomalous dispersion in the frequency range [between
maxon and roton].

It is evident from Fig. 1 that the lower the damping (or ohmic or scattering loss), the higher the gain in the
medium. The bandwidth of the laser amplifier is seen to becoming narrower with increasing gain. The concept of
bandwidth in the laser amplification is different from that in the band structure. Conventionally, the bandwidth
of an amplifier is defined as the full distance between the frequency (or energy) points at which the amplifier
gain has dropped to half the peak value. Another important issue is the nature of the electronic transitions: an amplifying (absorbing) transition implies a positive (negative) values of Im [$\epsilon^{-1}_{inter} (q,\omega)$]
and hence of $\alpha$($\omega$). Of course, one can always associate a suitable $\pm$ sign with $\alpha$($\omega$)
in order to give it the proper meaning for either amplifying or absorbing medium.

\begin{figure}[htbp]
\includegraphics*[width=8cm,height=9cm]{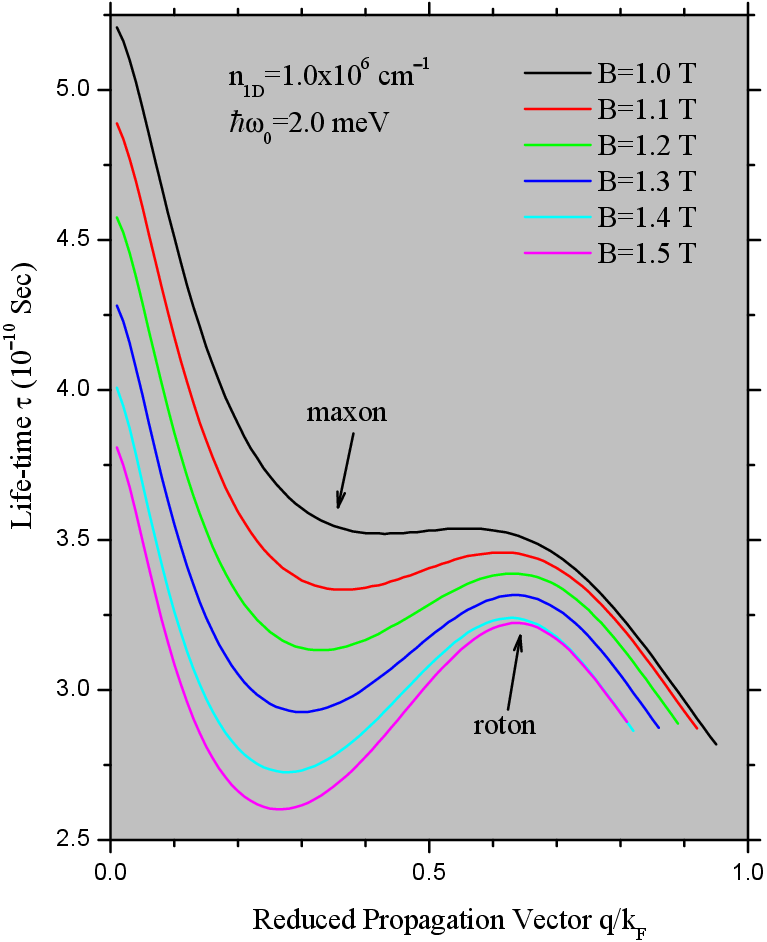}
\caption{(Color online) The life-time $\tau$ vs. the reduced propagation vector $q/k_F$ for the magnetoroton
for several values of the magnetic field $B$ [with $B \geq B_{th}$]. The other parameters are as given
inside the picture. We call attention to the trend: the stronger the magnetic field, the shorter the life-time,
and hence more susceptible the metastable state around the maxon.}
\label{fig2}
\end{figure}

One may argue that the sign specifying gain or loss should result from the calculus. As a matter of fact,
this is what we should expect if we are not sure of the characteristics of the medium. But if we know (as is the
case here) that NGV implies gain rather than loss in a certain frequency range [between maxon and roton], then we
do have the freedom to choose the sign of the damping term. Similar remarks as made on the sign convention here
can be seen in the biblical textbooks on lasers.

These characteristics open the possibility of designing MR-based electron device capable of amplifying a small
optical signal of definite wavelength. Figure 1 then clearly provides a platform for realizing the potential of
a magnetized quantum wire to act as an {\em active} laser medium. The situation is analogous to the (quasi-two dimensional) superlattices where the crystal can exhibit a negative resistance: it can refrain from consuming
energy like a resistor and instead feed energy into an oscillating circuit.


This paper reports on the device aspects of the magnetoroton excitations in the magnetized quantum wires in a
two-subband model within the framework of Bohm-Pines' full RPA. We have computed and discussed the gain
coefficient versus the excitation energy of the magnetoroton. This fundamental investigation suggests an
interesting, important, and significant application: the electronic device based on such magnetorotons can act
as an {\it active} laser medium. We believe that all the parameters involved in the process [such as the charge
density, magnetic field, confinement potential, propagation vector, etc.] that lead us to infer this proposal
are in the reach of the current technology. It is premature to predict whether the tantalizing concept of
magnetized quantum wire as an {\em active} laser medium will emerge as an exciting idea to be engaged in by the researchers. But certainly no other system of reduced dimensions has spoiled scientists and engineers with as
many appealing features to pursue.


\begin{acknowledgments}
During the course of this work the author has benefited from many enlightening discussions and communications
with some colleagues. I would like to particularly thank Naomi Halas and Peter Nordlander.
\end{acknowledgments}




\end{document}